\documentclass[12pt,a4paper]{article}
\usepackage{amsfonts}
\usepackage{amssymb}
\usepackage{amsmath}
\usepackage{latexsym}
\newtheorem{trm}{Theorem}
\newtheorem{lem}{Proposition}
\begin{document}

\begin{center}
{\Large \bf On no-go results for the nonlinear\\ \vspace{2mm}
Klein-Gordon-Maxwell equations} \\

\vspace{4mm}

M.N.~Smolyakov\\
\vspace{0.5cm} Skobeltsyn Institute of Nuclear Physics, Moscow
State University,
\\ 119991, Moscow, Russia\\
\end{center}

\begin{abstract}
In this paper we propose a new proof of some non-existence results
for the nonlinear Klein-Gordon-Maxwell system of equations. The
proof is based on the scaling arguments, i.e. special variations
of the fields, only. We also apply the obtained results to the
case of the simplest Q-balls and present some restrictions on
their existence.
\end{abstract}

\section{Introduction}
Nonlinear Klein-Gordon-Maxwell system of equations has been
extensively studied during the last years. There was obtained a
lot of existence and non-existence results, see, for example,
papers \cite{IDN}-\cite{IDN6} and references therein. A more
general case of Yang-Mills field coupled to a scalar field was
considered in \cite{GS} for a non-negative scalar filed potential
(see also review \cite{Malec}).

In this paper we will present a method for obtaining non-existence
results for the Klein-Gordon-Maxwell system of equations. This
method is based on the use of the scaling arguments proposed in
\cite{Derrick} and known as the Derrick theorem. Some of the
restrictions which can be obtained by our method coincide with
those obtained in \cite{DM}, and some of them are in agreement
with the restrictions presented in \cite{GS,Malec} for a more
general Yang-Mills-Klein-Gordon system. The difference between our
proof and the proofs of \cite{GS,DM} is that the proofs of
\cite{GS,DM} are based on some explicit properties of possible
solutions to the system of equations of motion, as well as on the
use of these equations itself, whereas our proof is based on the
arguments of \cite{Derrick}, which were applied to the system of
electromagnetic and scalar fields. It should also be mentioned
that analogous scaling arguments were discussed in \cite{DM} as an
alternative method which can be used at some step of that proof.
Our method is based on these scaling arguments, although we had to
use (only one) equation of motion in a special limiting case.

We will also apply the obtained results to the case of the
simplest Q-balls (a solitons in a system of a single complex
scalar field) and find that these results do not contradict the
known existence \cite{Coleman} and non-existence \cite{Strauss}
conditions and the existence of the solitons
\cite{Derrick2,Rosen}.

\section{The setup}
Let us consider the following form of the four-dimensional action:
\begin{equation}\label{act}
S=\int
d^{4}x\left[\eta^{\mu\nu}(D_{\mu}\varphi)^{*}D_{\nu}\varphi-m^{2}\varphi^{*}\varphi-V(\varphi^{*}\varphi)-\frac{1}{4}F^{\mu\nu}F_{\mu\nu}\right],
\end{equation}
where $\eta_{\mu\nu}=diag(1,-1,-1,-1)$ is the flat Minkowski
metric, $D_{\mu}\varphi=\partial_{\mu}\varphi-ieA_{\mu}\varphi$.
Of course, the term $m^{2}\varphi^{*}\varphi$ can be incorporated
into $V(\varphi^{*}\varphi)$, but we will retain it in order to
correspond to the notations used in \cite{DM}. Note that in papers
\cite{GS,Malec} the mass term is incorporated into
$V(\varphi^{*}\varphi)$.

In what follows we will focus on the standing wave solutions
\begin{equation}\label{sol1}
\varphi(t,\vec x)=e^{i\omega t}\phi(\vec x)
\end{equation}
with the real function $\phi(\vec x)$. We also suppose that
\begin{equation}\label{sol2}
A_{i}\equiv 0,\quad A_{0}(t,\vec x)=A_{0}(\vec x)
\end{equation}
for $i=1,2,3$. In this case we can use the following
three-dimensional effective action with
$V(\varphi^{*}\varphi)=V(\phi)$ instead of (\ref{act})

\begin{eqnarray}\label{act2}
S=\int
d^{3}x\left[-\partial_{i}\phi\partial_{i}\phi+\frac{1}{2}\partial_{i}A_{0}\partial_{i}A_{0}+
(\omega^2-m^2)\phi^2-\right.\\ \nonumber\left. -2e\omega
A_{0}\phi^{2}+e^2A_{0}^2\phi^{2}-V(\phi)\right],
\end{eqnarray}
where
$\partial_{i}\phi\partial_{i}\phi=\sum\limits_{k=1}^{3}\partial_{k}\phi\partial_{k}\phi$.
We will be looking for non-topological solitons such that
\begin{eqnarray}\label{finite}
\int d^3 x\phi^{2}<\infty,\quad \int d^3
x\partial_{i}\phi\partial_{i}\phi<\infty, \\ \nonumber \left|\int
d^3x V(\phi)\right|<\infty,\quad \left|\int d^3x
\frac{dV(\phi)}{d\phi}\phi\right|<\infty,\\ \nonumber \int d^3
x\partial_{i}A_{0}\partial_{i}A_{0}<\infty,\quad
\lim\limits_{x^{i}\to\pm\infty}A_{0}=0,\quad
\lim\limits_{x^{i}\to\pm\infty}\phi=0.
\end{eqnarray}
It is straightforward to get from (\ref{act2}) the corresponding
equations of motion for the fields $\phi$, $A_{0}$, they can be
found, for example, in \cite{DM}. We do not present these
equations here to stress that we will not use them in their
explicit form to obtain the no-go results.

Let ${\phi(\vec x), A_{0}(\vec x)}$ be a localized solution to the
equations of motion following from action (\ref{act2}). It means
that
\begin{equation}\label{ds}
\delta S(\phi, A_{0})=0.
\end{equation}
Let us denote
\begin{eqnarray}\label{def}
\int d^3 x\,\phi^{2}=V_{1}\ge 0,\quad \int d^3
x\,\partial_{i}\phi\partial_{i}\phi=\Pi_{1}\ge 0,\\ \nonumber
\frac{1}{2}\int d^3
x\,\partial_{i}A_{0}\partial_{i}A_{0}=\Pi_{2}\ge 0,\quad e^2\int
d^3 x\,A_{0}^2\phi^{2}=I_{2}\ge 0,\\
\nonumber 2e\omega \int d^3 x\,A_{0}\phi^{2}=I_{1},\quad \int d^3
x\,V(\phi)=V_{2}.
\end{eqnarray}
Now we can proceed to specific examples.

\section{No-go results}
\subsection{Klein-Gordon-Maxwell system}
First, let us consider the potential $V(\phi)$ to have the form
\begin{equation}\label{vpotent}
V(\phi)=\gamma\left(\phi^{2}\right)^{p/2}=\gamma|\phi|^{p}
\end{equation}
with $p>1$, which is often used for examining the
Klein-Gordon-Maxwell system of equations. In this case
\begin{eqnarray}
V_{2}=\int d^3 x\,V(\phi)=\gamma\int d^3 x\,|\phi|^{p}=\gamma
\tilde V_{2},
\end{eqnarray}
where $\tilde V_{2}\ge 0$. The potential of form (\ref{vpotent})
allows one to see how the generalized scale change method works in
a simple case.

Let us consider the following modification of the solution ${\phi,
A_{0}}$:
\begin{eqnarray}\label{resc1}
\phi(\vec x)\to\lambda^{\alpha}\phi(\lambda \vec x),\\
\label{resc2} A_{0}(\vec x)\to\lambda^{\beta}A_{0}(\lambda \vec
x).
\end{eqnarray}
Since $\lambda=1$ corresponds to the solution to (\ref{ds}), the
following identity holds
\begin{equation}\label{ds1}
\frac{dS(\lambda)}{d\lambda}\biggl|_{\lambda=1}\biggr.=0.
\end{equation}
Substituting (\ref{resc1}) and (\ref{resc2}) into (\ref{act2}) and
using (\ref{def}), (\ref{ds1}) we get
\begin{eqnarray}\label{ds2}
-(2\alpha-1)\Pi_{1}
+(2\beta-1)\Pi_{2}+(2\alpha-3)(\omega^2-m^2)V_{1}-\\
\nonumber
-(2\alpha+\beta-3)I_{1}+(2\alpha+2\beta-3)I_{2}-(p\alpha-3)\gamma\tilde
V_{2}=0.
\end{eqnarray}
The term $I_{1}$ is not of a definite sign, so below we will
consider the case $\beta=3-2\alpha$. Thus we obtain from
(\ref{ds2})
\begin{eqnarray}\label{ds3}
-(2\alpha-1)\Pi_{1}
+(5-4\alpha)\Pi_{2}+(2\alpha-3)(\omega^2-m^2)V_{1}+\\
\nonumber +(3-2\alpha)I_{2}-(p\alpha-3)\gamma\tilde V_{2}=0.
\end{eqnarray}
Now we are ready to discuss restrictions on the existence of
solitons coming from (\ref{ds3}) for different values of $\alpha$,
$\gamma$ and $p$.
\begin{lem}
For the potential of form (\ref{vpotent}) non-topological solitons
of form (\ref{sol1}), (\ref{sol2}), (\ref{finite}) are absent if
\begin{enumerate}
\item $\gamma=0$.
\item $\gamma>0$
\begin{itemize}
\item $p\ge 2$,
\item $1<p<2$ {\rm and} $m^2\ge \omega^2$.
\end{itemize}
\item $\gamma<0$
\begin{itemize}
\item
$1<p\le 2$,
\item
$p\ge 6$ {\rm and} $m^2\ge \omega^2>0$.
\end{itemize}
\end{enumerate}
\end{lem}
{\bf Remark:} the absence of solitons in the case $\gamma=0$ was
shown in \cite{DM} (the absence of spherically symmetric solitons
in this case was shown in \cite{Deumens}); restrictions for the
case $\gamma>0$ are in agreement with those presented in \cite{GS}
for a more general case of Yang-Mills-Klein-Gordon system (they
are $p\ge 4$ for $m=0$ and $p>2$ for $m\ne 0$ in our notations);
restrictions for the case $\gamma< 0$ coincide with those obtained
in paper \cite{DM}.
\newpage

{\bf Proof:}
\begin{enumerate}
\item $\gamma=0$.\\
In this case it is convenient to take $\alpha=\frac{3}{2}$.
Equation (\ref{ds3}) takes the form
$$-2\Pi_{1}-\Pi_{2}=0.$$
Since by definition (\ref{def}) $\Pi_{1}\ge 0$, $\Pi_{2}\ge 0$,
the latter identity implies $\Pi_{1}=\Pi_{2}=0$ and thus
$\phi=A_{0}\equiv 0$. There are no solitons of form (\ref{sol1}),
(\ref{sol2}), (\ref{finite}) for $\gamma=0$.

Analogous considerations will be used below for the other values
of $\gamma$.
\item $\gamma>0$.\\
In this case it is also convenient to take $\alpha=\frac{3}{2}$.
Equation (\ref{ds3}) takes the form
$$-2\Pi_{1}-\Pi_{2}-3\left(\frac{p}{2}-1\right)\gamma\tilde
V_{2}=0.$$ Since $\tilde V_{2}\ge 0$, it implies that for $p\ge 2$
$\phi=A_{0}\equiv 0$ and there are no solitons of form
(\ref{sol1}), (\ref{sol2}), (\ref{finite}).

Now let us take $\alpha>\frac{3}{2}$. Identity (\ref{ds3}) takes
the form
$$
(2\alpha-1)\Pi_{1}
+(4\alpha-5)\Pi_{2}+(2\alpha-3)(m^2-\omega^2)V_{1}+(2\alpha-3)I_{2}+(p\alpha-3)\gamma\tilde
V_{2}=0.
$$
In this case if $m^2\ge \omega^2$ and $p\alpha\ge 3$ then
$\phi=A_{0}\equiv 0$. Considering $\alpha>\frac{3}{2}$ and
$p\alpha\ge 3$ together one can conclude that for any $p>1$ there
exist $\alpha: \alpha>\frac{3}{2}$ and $p\alpha\ge 3$ hold
simultaneously. Thus, for $m^2\ge \omega^2$ there are no solitons
of form (\ref{sol1}), (\ref{sol2}), (\ref{finite}).

Considering other values of the parameter $\alpha$ does not
provide any additional restrictions on the existence of solitons.
\item $\gamma<0$ (this case was considered in \cite{DM}).\\
First let us again take $\alpha=\frac{3}{2}$. Equation (\ref{ds3})
again takes the form
$$-2\Pi_{1}-\Pi_{2}-3\left(\frac{p}{2}-1\right)\gamma\tilde
V_{2}=0.$$ But since $\gamma<0$, it implies that now
$\phi=A_{0}\equiv 0$ and there are no solitons of form
(\ref{sol1}), (\ref{sol2}), (\ref{finite}) if $p\le 2$.

It is also useful to consider the case $\alpha=\frac{1}{2}$.
Identity (\ref{ds3}) takes the form
\begin{equation}\label{p6}
3\Pi_{2}-2(\omega^2-m^2)V_{1}+2I_{2}-\left(\frac{p}{2}-3\right)\gamma\tilde
V_{2}=0.
\end{equation}
Now $\phi=A_{0}\equiv 0$ and there are no solitons of form
(\ref{sol1}), (\ref{sol2}), (\ref{finite}) if $p\ge 6$ and
$m^2>\omega^2$ or $p>6$ and $m^2\ge\omega^2$. The case $p=6$ and
$m^2=\omega^2\ne 0$ should be considered separately. Equation
(\ref{p6}) gives us only the condition $A_{0}\equiv 0$ for $p=6$
and $m^2=\omega^2$. Nevertheless, it is easy to see that
$\phi\equiv 0$ also holds. Indeed, equation of motion for the
field $A_{0}$ coming from (\ref{act2}) takes the form
$\phi^2\equiv 0$ for $A_{0}\equiv 0$, $\omega\ne 0$ and,
consequently, $\phi\equiv 0$ whenever $A_{0}\equiv 0$ and
$\omega\ne 0$. The latter statement is quite obvious from the
physical point of view, because only the solution with the zero
charge density can provide the absence of the electric field in
the whole space.

Considering other values of the parameter $\alpha$ also does not
provide any additional restrictions on the existence of solitons.
\hspace{9.2cm} $\Box$
\end{enumerate}

It should be noted that an analogous generalized rescaling of the
field $\phi$ (but with $\alpha=1$) was discussed in \cite{DM} as a
possible alternative method which can be used at some step of the
proof presented in \cite{DM}.

The method of generalized rescaling presented above can be used in
more general cases. Let us consider quite a general form of the
potential $V(\phi)$ such that
\begin{equation}\label{vpotgeneral}
V(\phi)|_{\phi=0}=0,\quad \frac{dV(\phi)}{d\phi}|_{\phi=0}=0.
\end{equation}
The latter condition ensures that the trivial solution is
$\phi\equiv 0$, $A_{0}\equiv 0$. Using (\ref{resc1}) one can get
\begin{eqnarray}
V_{2}(\lambda)&=&\int d^3 x\,V(\lambda^{\alpha}\phi(\lambda
x))=\frac{1}{\lambda^3}\int d^3 \tilde
x\,V(\lambda^{\alpha}\phi(\tilde x))=\\
\nonumber &=&\frac{1}{\lambda^3}\int d^3
x\,V(\lambda^{\alpha}\phi(x)),
\end{eqnarray}
where $V_{2}(1)=V_{2}$, and
\begin{eqnarray}
\frac{dV_{2}(\lambda)}{d\lambda}\biggl|_{\lambda=1}\biggr.=\alpha\int
d^3 x\,\left(\frac{dV(\phi)}{d\phi}\phi(x)\right)-3V_{2}.
\end{eqnarray}
Equation (\ref{ds3}) transforms into
\begin{eqnarray}\label{ds4}
-(2\alpha-1)\Pi_{1}
+(5-4\alpha)\Pi_{2}+(2\alpha-3)(\omega^2-m^2)V_{1}+\\
\nonumber (3-2\alpha)I_{2}-\left(\alpha\int d^3
x\,\left(\frac{dV(\phi)}{d\phi}\phi(x)\right)-3V_{2}\right)=0.
\end{eqnarray}
Now we are ready to obtain more general non-existence results.

\begin{trm}
For the potential of form (\ref{vpotgeneral}) non-topological
solitons of form (\ref{sol1}), (\ref{sol2}), (\ref{finite}) are
absent at least if one of the following inequalities fulfills
\begin{itemize}
\item
$\frac{dV(\phi)}{d\phi}\phi-2V(\phi)\ge 0,$
\item
$4(m^2-\omega^2)\phi^{2}\ge
\frac{dV(\phi)}{d\phi}\phi-6V(\phi),$\quad $\omega\ne 0$,
\item
$V(\phi)-(\omega^2-m^2)\phi^2\ge 0$
\end{itemize}
for any $\phi$ (or at least for that range of values of the field
$\phi$ which is supposed for a solution).
\end{trm}
{\bf Remark:} The first two relations coincide with those obtained
in \cite{DM} (up to the notations). The first relation also
coincides with the restriction presented in \cite{GS} for a more
general case of Yang-Mills-Klein-Gordon system with $m\ne 0$ and
$m^2\phi^2+V(\phi)\ge 0$ (in our notations). The third relation
for the static case $\omega=0$ and for $m=0$ simply gives the
Derrick theorem \cite{Derrick}.

{\bf Proof:}\\ First we take $\alpha=\frac{3}{2}$. Equation
(\ref{ds4}) takes the form
\begin{eqnarray}\label{ds5}
-2\Pi_{1}-\Pi_{2}-\int d^3
x\frac{3}{2}\left(\frac{dV(\phi)}{d\phi}\phi(x)-2V(\phi)\right)=0.
\end{eqnarray}
So if $\frac{dV(\phi)}{d\phi}\phi-2V(\phi)\ge 0$ for any $\phi$
(or at least for that range of values of the field $\phi$ which is
supposed for a solution), then solitons of form (\ref{sol1}),
(\ref{sol2}), (\ref{finite}) are absent in the theory (again
because (\ref{ds5}) implies that $\Pi_{1}=\Pi_{2}=\int d^3
x\left(\frac{dV(\phi)}{d\phi}\phi(x)-2V(\phi)\right)=0$). The
latter inequality can be rewritten as
\begin{equation}\label{Qno}
\phi^{3}\frac{d\left(\frac{V(\phi)}{\phi^{2}}\right)}{d\phi}\ge 0.
\end{equation}

Second we take $\alpha=\frac{1}{2}$. Identity (\ref{ds4}) takes
the form
\begin{eqnarray}\label{ds6}
3\Pi_{2}-2(\omega^2-m^2)V_{1}+2I_{2}-\left(\frac{1}{2}\int
\frac{dV(\phi)}{d\phi}\phi(x)d^3 x-3V_{2}\right)=0.
\end{eqnarray}
Using $\int d^3 x\,\phi^{2}=V_{1}$ one easily realizes that
$A_{0}\equiv 0$ if
\begin{equation}
4(m^2-\omega^2)\phi^{2}\ge \frac{dV(\phi)}{d\phi}\phi-6V(\phi)
\end{equation}
for any $\phi$. Using the equation of motion for the field $A_{0}$
we can also get $\phi\equiv 0$ provided $\omega\ne 0$ (see the
paragraph just after equation (\ref{p6})).

Now let us take $\alpha=0$. Equation (\ref{ds4}) takes the form
\begin{eqnarray}\label{ds55}
\Pi_{1}+5\Pi_{2}-3(\omega^2-m^2)V_{1}+3I_{2}+3V_{2}=0,
\end{eqnarray}
from which we get that solitons of form (\ref{sol1}),
(\ref{sol2}), (\ref{finite}) are absent in the theory if
\begin{equation}\label{vminom}
V(\phi)-(\omega^2-m^2)\phi^2\ge 0.
\end{equation}

Now let us consider transformations of form
\begin{eqnarray}\label{noresc1}
\phi(\vec x)\to\lambda^{\alpha}\phi(\vec x),\\
\label{noresc2} A_{0}(\vec x)\to\lambda^{\beta}A_{0}(\vec x),
\end{eqnarray}
for the system with action (\ref{act2}), i.e. transformations
without rescaling of the coordinates. With $\beta=-2\alpha$ (this
choice vanishes the term $\sim I_{1}$) we get from (\ref{ds1}) for
any $\alpha\ne 0$
\begin{equation}\label{noresc3}
-2\Pi_{1}-4\Pi_{2}+2(\omega^2-m^2)V_{1}-2I_{2}-\int d^3
x\,\frac{dV(\phi)}{d\phi}\phi(x)=0,
\end{equation}
which leads to a new additional restriction: solitons are absent
if
\begin{equation}\label{lastrestr}
\frac{dV(\phi)}{d\phi}\phi-2(\omega^2-m^2)\phi^{2}\ge 0.
\end{equation}
This relation for the static case $\omega=0$ is in agreement with
that presented in \cite{Malec} for the static
Yang-Mills-Klein-Gordon system. But since
$V(\phi)-(\omega^2-m^2)\phi^2|_{\phi=0}=0$ (see
(\ref{vpotgeneral})), from inequality (\ref{lastrestr})
automatically follows inequality (\ref{vminom}).\hspace{7cm}
$\Box$

It should be noted that all the restrictions presented in
Proposition~1 follow from Theorem~1. But the restrictions
presented in Theorem~1 are more general, because they deal not
only with the parameters of the potential ($\gamma$, $m$, $p$) and
frequency $\omega$, but with the values of the scalar field
potential and values of its first derivative for all values of the
field $\phi$ (or at least for that range of values of the scalar
field which is supposed for a solution).

\subsection{Q-balls}
The arguments presented above can be easily applied to the case of
the simplest Q-balls \cite{Coleman,Rubakov}, which are solitary
wave solutions in the case of absence of electromagnetic field (we
also set $m=0$ in this case).
\begin{lem}
For the potential of form (\ref{vpotgeneral}) Q-ball solutions of
form (\ref{sol1}) and such that
$\lim\limits_{x^{i}\to\pm\infty}\phi=0$ are absent at least if one
of the following inequalities fulfills
\begin{itemize}
\item
$\frac{dV(\phi)}{d\phi}\phi-2V(\phi)\ge 0,$
\item
$4\omega^2\phi^{2}+\frac{dV(\phi)}{d\phi}\phi-6V(\phi)<0$\qquad
{\rm or}\qquad
$4\omega^2\phi^{2}+\frac{dV(\phi)}{d\phi}\phi-6V(\phi)>0$,
\item
$V(\phi)-\omega^2\phi^2\ge 0$
\end{itemize}
for any $\phi\ne 0$\, (or at least for that range of values of the
field $\phi$ which is supposed for a solution).
\end{lem}
{\bf Remark:} The second and third relations coincide with those
obtained in \cite{Strauss} (see Theorem~1 in \cite{Strauss} in a
three-dimensional case).

{\bf Proof:}\\
The proof directly follows from equations (\ref{ds5}), (\ref{ds6})
and (\ref{ds55}) if one simply takes $\Pi_{2}=I_{1}=I_{2}\equiv 0$
and $m=0$. The difference is only in the case of item~2. The
situation differs from that of the Klein-Gordon-Maxwell system
because we have no equation for the field $A_{0}$ which can
provide additional constraints in the limiting case
$4\omega^2\phi^{2}+\frac{dV(\phi)}{d\phi}\phi-6V(\phi)=0$ (see eq.
(\ref{ds6})). The absence of terms $\Pi_{2}$ and $I_{2}$ also
leads to the existence of two inequalities instead of one
(equation (\ref{ds6}) with $\Pi_{2}=I_{1}=I_{2}\equiv 0$ and $m=0$
does not hold if
$4\omega^2\phi^{2}+\frac{dV(\phi)}{d\phi}\phi-6V(\phi)>0$ or
$4\omega^2\phi^{2}+\frac{dV(\phi)}{d\phi}\phi-6V(\phi)<0$).
\hspace{10.5cm} $\Box$

It is illustrative to apply these restrictions to the Q-ball
solution proposed in \cite{Coleman}. It was shown in
\cite{Coleman} that a Q-ball can exist if
$\frac{V(\phi)}{\phi^{2}}$ has a minimum at $\phi_{0}\ne 0$. For
simplicity suppose that $\phi_{0}>0$. The existence of a minimum
implies that
$\frac{d\left(\frac{V(\phi)}{\phi^{2}}\right)}{d\phi}$ changes its
sign at $\phi=\phi_{0}$ and condition (\ref{Qno}) is not
fulfilled. In this case in principle $\int d^3
x\left(\frac{dV(\phi)}{d\phi}\phi(x)-2V(\phi)\right)$ can be
negative and there are no restrictions on the existence of
solitons. Thus, the results described above do not contradict
those obtained in \cite{Coleman}.

Other examples of solutions, for which we can check our results,
can be found in \cite{Derrick2,Rosen}. Solution in \cite{Derrick2}
corresponds to $V(\phi)=\mu^2\phi^2-g^2\phi^4$ and was found
numerically, whereas solution in \cite{Rosen} corresponds to
$V(\phi)=\mu^2\phi^2-g\phi^2\ln(\phi^2)$ and was found
analytically. Although the first solution was found numerically
and appeared to be unstable, it can be easily shown by using its
explicit form (see \cite{Derrick2}) that all the non-existence
conditions for the Q-balls presented above (four conditions)
appear to be violated by this solution. The same analysis can be
made for the "at rest" solution of \cite{Rosen} with $g>0$, in
this case all the non-existence conditions also appear to be
violated by the solution. Thus, the presented results again do not
contradict the existence of the solitons.

\section*{Acknowledgements}

The author is grateful to I.P.~Volobuev for discussions. The work
was supported by grant of the "Dynasty" Foundation, FASI state
contract 02.740.11.0244 and RFBR grant 08-02-92499-CNRSL$\_$a.

\end{document}